\documentclass{aa}  

\usepackage{graphicx}
\usepackage{xcolor}
\usepackage{multirow}
\usepackage{array}
\usepackage{txfonts}
\usepackage{hyperref}
\addto\extrasenglish{

}

\begin{document} 

   \title{Characterization of carbon dioxide on Ganymede and Europa supported by experiments: Effects of temperature, porosity, and mixing with water}

   \author{L. Schiltz
          \inst{1}
          \and
          B. Escribano
          \inst{2}
          \and
           G. M. Mu\~noz Caro
          \inst{2}
          \and
          S. Cazaux
          \inst{1,3}
          \and
          C. del Burgo Olivares\inst{2}
          \and
          H. Carrascosa\inst{2} 
          \and \\
           I. Boszhuizen\inst{1}
           \and
          C. Gonz\'alez D\'iaz\inst{2}
          \and
          Y.-J. Chen \inst{4}
          \and
          B. M. Giuliano \inst{5}
          \and
          P. Caselli\inst{5} 
          }

   \institute{Faculty of Aerospace Engineering, Delft University of Technology, Delft, The Netherlands\\
              \email{s.m.cazaux@tudelft.nl}
         \and
         Centro de Astrobiolog\'{\i}a (CSIC-INTA), Ctra. de Ajalvir, km 4, Torrej\'on de Ardoz, 28850 Madrid, Spain
         \and
         Leiden Observatory, Leiden University, P.O. Box 9513, NL 2300 RA Leiden, The Netherlands
         \and
         Department of Physics, National Central University, Jhongli City, Taoyuan County 32054, Taiwan 
         \and
         Max Planck Institute for Extraterrestrial Physics, Postfach 1312, 85741 Garching, Germany
             }

   \date{Received XXXXXX; accepted XXXX}

 
  \abstract
   {The surfaces of icy moons are primarily composed of water ice that can be mixed with other compounds, such as carbon dioxide. The carbon dioxide (CO$_2)$ stretching fundamental band observed on Europa and Ganymede appears to be a combination of several bands that are shifting location from one moon to another.}
   {We investigate the cause of the observed shift in the CO$_2$ stretching absorption band experimentally. We also explore the spectral behaviour of CO$_2$ ice by varying the temperature and concentration.} 
   {We analyzed pure CO$_2$ ice and ice mixtures deposited at 10 K under ultra-high vacuum conditions using Fourier-transform infrared (FTIR) spectroscopy and temperature programmed desorption (TPD) experiments. Laboratory ice spectra were compared to JWST observation of Europa's and Ganymede's leading hemispheres. The simulated IR spectra were calculated using density functional theory (DFT) methods, exploring the effect of porosity in CO$_2$ ice.}
   {Pure CO$_2$ and CO$_2$-water ice show distinct spectral changes and desorption behaviours at different temperatures, revealing intricate CO$_2$ and H$_2$O interactions. The number of discernible peaks increases from two in pure CO$_2$ to three in CO$_2$-water mixtures.}
   {The different CO$_2$ bands were assigned to $\tilde{\nu}_{3,1}$ (2351 cm$^{-1}$, 4.25 $\mu$m) caused by CO$_2$ dangling bonds (CO$_2$ found in pores or cracks) and $\tilde{\nu}_{3,2}$ (2345 cm$^{-1}$, 4.26 $\mu$m) due to CO$_2$ segregated in water ice, whereas $\tilde{\nu}_{3,3}$ (2341 cm$^{-1}$, 4.27 $\mu$m) is due to CO$_2$ molecules embedded in  water ice. The JWST NIRSpec CO$_2$ spectra for Ganymede and for Europa can be fitted with two Gaussians attributed to $\tilde{\nu}_{3,1}$ and $\tilde{\nu}_{3,3}$. For Europa, $\tilde{\nu}_{3,1}$ is located at lower wavelengths due to a lower temperature. The Ganymede data reveal latitudinal variations in CO$_2$ bands, with $\tilde{\nu}_{3,3}$ dominating in the pole and $\tilde{\nu}_{3,1}$ prevalent in other regions. This shows that CO$_2$ is embedded in water ice at the poles and it is present in pores or cracks in other regions. Ganymede longitudinal spectra reveal an increase of the CO$_2$ $\tilde{\nu}_{3,1}$ band throughout the day, possibly due to ice cracks or pores caused by large temperature fluctuations.}
    
   \keywords{Carbon dioxide ices -- methods: laboratory --
                JWST observations --
                infrared spectroscopy -- icy moons 
               }
\titlerunning{CO$_2$ on icy moons} 
   \maketitle
%

\section{Introduction}
Mixed ices of CO$_2$ and water are known to be present in the frozen nuclei of comets \citep{crovisier_a, crovisier_b} in various satellites of our Solar System \citep[e.g.][]{grundy,buratti} and as components of dust interstellar particles \citep[e.g][]{dartois,draine}. On Ganymede, the CO$_2$ stretching fundamental band, which is ordinarily at 2341 cm$^{-1}$ (4.27 $\mu$m) \citep{Falk1987}, has been observed with the Near Infrared Mapping Spectrometer (NIMS) aboard the Galileo spacecraft \citep{Mccord1998}. The location of this absorption band is displaced slightly to a shorter wavelength (2345 cm$^{-1}$, 4.26 $\mu$m) indicating that the CO$_2$ is bound, or trapped, in a host. Recent \textit{James Webb} Space Telescope (JWST) observations \citep{bockelee} have shown variations in the latitude and longitude  of the CO$_2$ bands on Ganymede's surface. In the boreal region of the leading hemisphere, the CO$_2$ band is dominated by the 2341 cm$^{-1}$ (4.27 $\mu$m) band, consistent with CO$_2$ trapped in amorphous water ice, while at equatorial latitudes (and especially on dark terrains) the observed band is broader and located around 2345 cm$^{-1}$ (4.26 $\mu$m), suggesting CO$_2$ adsorbed on non-icy materials, such as minerals or salts. On Europa, CO$_2$ was detected with the Galileo/NIMS instrument at 2351 cm$^{-1}$ (4.25 $\mu$m) and mostly located on the anti-Jovian and trailing sides \citep{Hansen2008}; however, observations in the near-infrared (NIR) were not able to confirm the presence of CO$_2$ \citep{Mishra2021}. Recent JWST observations by \cite{villanueva} have highlighted the presence of CO$_2$ on Europa's surface, with bands located at 2351 cm$^{-1}$ (4.25 $\mu$m) and 2341 cm$^{-1}$ (4.27 $\mu$m). These observations suggest that CO$_2$ is mixed with other compounds and that carbon is sourced from within Europa, probably from the liquid sub-surface ocean.

Prompted by these findings, solid H$_2$O/CO$_2$ ice mixtures, as laboratory analogues of astrophysical objects, have been studied for many years, mainly by infrared spectroscopy and mass spectrometry, in controlled warm-up  experiments \citep[i.e.][]{sandford,ehrenfreund,bernstein,kumi,malyk}. From these studies, the interaction between CO$_2$ and H$_2$O on a molecular level has been shown to cause significant changes in the position and profile of CO$_2$ peaks in the IR. However, a concise answer to why the CO$_2$ stretching fundamental band on icy moons, like Europa and Ganymede, shows frequency shifts remains elusive.

In the present work, we attempt to find an explanation for the different CO$_2$ bands to characterise the carbon dioxide on Europa and Ganymede. This is achieved by investigating the influence of temperature on the spectral properties of laboratory CO$_2$ ice, and in particular the behaviour of the CO$_2$ stretching mode, which (in its pure form) peaks around 2344 cm$^{-1}$ (4.27 $\mu$m). We note that CO$_2$ is first studied in its pure form and afterwards co-deposited with water so that the H$_2$O:CO$_2$ deposition ratio is varied. In particular, we considered low (<5$\%$) and high (>25$\%$) carbon dioxide concentrations in the ice. Experiments were carried out at temperatures ranging between 10 K and 160 K and ultra high vacuum conditions. This research uses a combination of the following two techniques: temperature programmed desorption (TPD) and Fourier-transform infrared spectroscopy (FTIR) in transmittance, across a variety of experimental conditions. Additionally, Gaussian deconvolution was applied to the spectra obtained experimentally. This technique was first applied to pure CO$_2$ ice and afterwards to the ice mixture H$_2$O:CO$_2$. It was found that deconvolution of the asymmetric stretching band $\nu_3$   requires two Gaussians for pure CO$_2$ and three when mixed with water. The evolution of the integrated absorbance areas, band shape and position as a function of temperature can be compared to the TPD curve in order to assign the bands to different molecular interactions, thus characterising the type of CO$_2$ ice. The synergy between laboratory studies and data gathered by JWST enriches our comprehensive approach to understanding CO$_2$ behaviour on icy worlds, enabling a thorough exploration of the fundamental properties of CO$_2$ ice in various conditions.

The structure of this paper is organized as follows. In \autoref{Ch 2}, we outline the experimental setup and the determination of the gas mixture ratio. In \autoref{ch3}, we focus on the spectroscopic behaviour of pure CO$_2$ ice, elucidating the shape and location of the $\nu_3$ stretching band, alongside the TPD curve. In \autoref{ch4} and \autoref{ch5}, we describe the acquired laboratory spectra for low and high CO$_2$ concentrations in water ice as a function of temperature, accompanied by explanations of integrated absorbance areas and TPD curves.
In \autoref{sec:discussion} and \autoref{sec:DFT}, we discuss the results and assign the IR vibrational modes from experiments and simulations.
Finally, \autoref{ch7} delves into the application of findings to icy moons, specifically Europa and Ganymede, involving a comparison of new experimental results with recent observations.

\section{Experimental method}
\label{Ch 2}

\subsection{ISAC set-up}
\begin{figure}[ht]
        \centering
        \includegraphics[width=\hsize]{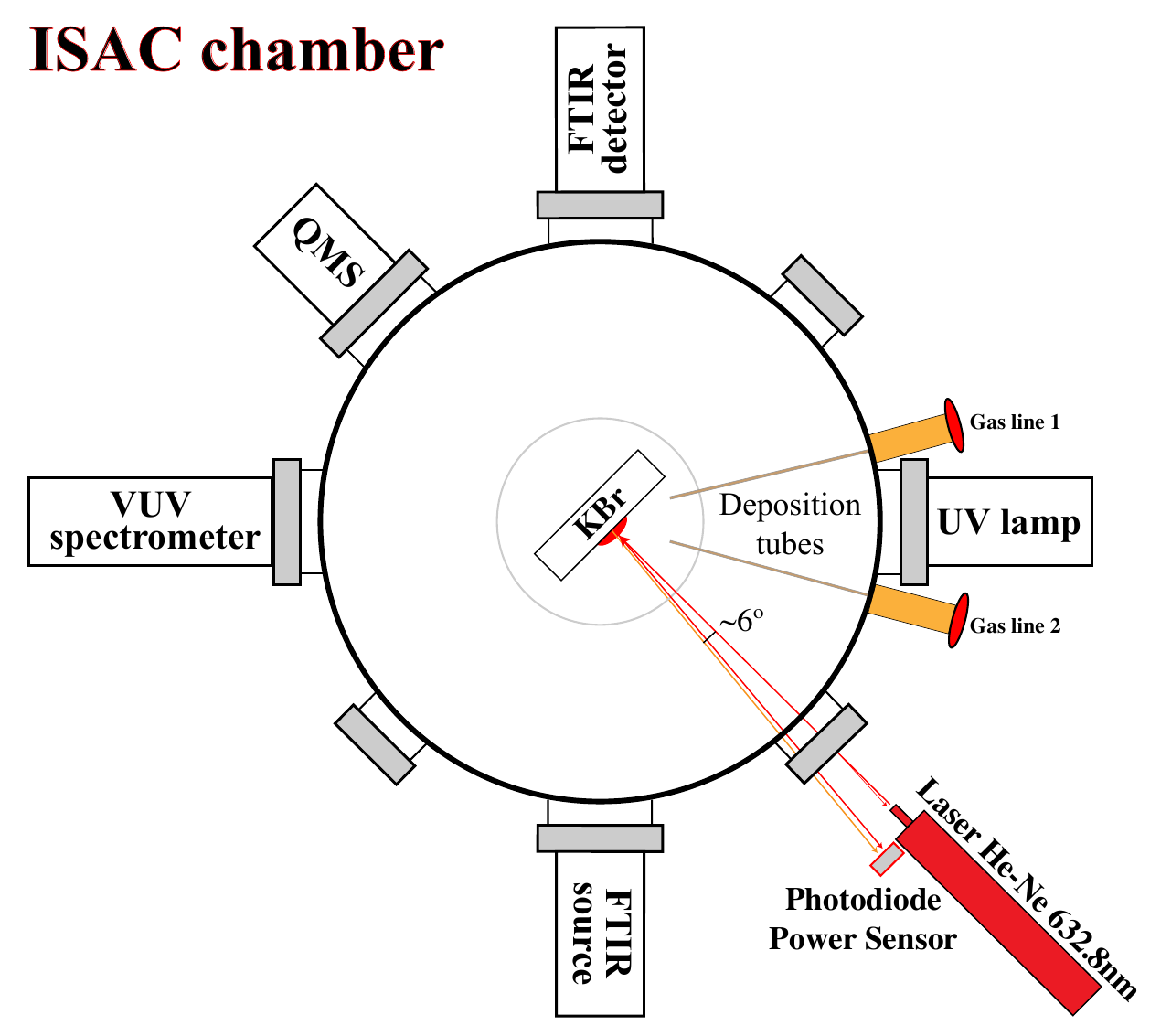}
        \caption{ISAC cross-section at the ice sample level. The measuring devices and sensors, including the laser interferometry, are displayed. The FTIR source is positioned on one side and the FTIR detector on the opposite one. The UV spectrometer is placed directly across from the vacuum-UV lamp, see \cite{gonzalez} for more details.}
        \label{fig:ISAC setup}
    \end{figure}

\indent The experiments reported in this work have been performed with the Interstellar Astrochemistry Chamber (ISAC), described in more detail in \cite{munoz}. ISAC is an ultra-high vacuum (UHV) chamber with a base pressure of 4 $\times$ 10$^{-11}$ mbar, designed to simulate the conditions present in the interstellar medium (ISM), regarding temperature, pressure and ultraviolet (UV) radiation field. A closed-cycle He cryostat allows for the tip of the cold finger to cool down to 8 K, where a sample holder with a potassium bromide (KBr) window acting as the substrate for ice deposition is located. A schematic representation is shown in \autoref{fig:ISAC setup}. A Lakeshore temperature controller 331 with 0.1 K accuracy was used. The gas line, which permits the introduction of gas species with regulated compositions, is linked to the main chamber through a leak valve. The identification of the species present in the chamber is done via a quadrupole mass spectrometer (QMS, Pfeiffer Vacuum, Prisma QMS 200). The valve is opened during deposition, and the gas is delivered to the cold substrate through a deposition tube. This tube's end is roughly placed at a distance of 3 cm from the substrate. FTIR in transmittance mode using a Bruker Vertex 70 at a working spectral resolution of 2 cm$^{-1}$ is used to record infrared spectra during deposition, and later during the TPD using a heating ramp in K/min. ISAC uses laser interferometry at 632.8 nm to assess changes in ice thickness (\cite{gonzalez}). A 5.0 mW He-Ne red laser and a Silicon Photodiode Power Sensor to measure the optical power of the laser light (model S120C) are mounted at an angle of 6$^{\circ}$, as seen in \autoref{fig:ISAC setup}.

In general, ices were grown at 10 K by opening the leak valve of gas line 2 in \autoref{fig:ISAC setup}, feeding a mixture of H$_2$O and CO$_2$ into the main chamber, at a deposition pressure of 2$\times$10$^{-7}$ mbar. To investigate the influence of deposition temperature and pressure on the spectral properties of CO$_2$ ice (in particular, the behaviour of the CO$_2$ stretching mode), some experiments were also carried out at 70 K and higher deposition pressure. 
The experiments presented in  \autoref{table:1} were repeated for reproducibility or for suspected contamination.
During the deposition of the gas mixture to form the ice, infrared spectra at 45$^{\circ}$ incidence of the beam relative to the substrate were collected every 300 s. Warming up the samples after deposition was done at 0.2 K min$^{-1}$ until a temperature of 210 K was attained, ensuring that both the CO$_2$ and H$_2$O were entirely desorbed. 
IR measurements were taken every 300 s again, yielding a spectrum every 1 K temperature increment.

The key parameters measured during the experiments using various devices include: the laser intensity to measure the ice thickness \citep{gonzalez},  pressure inside the chamber, the ion currents of the relevant molecules (via QMS), and temperature. The temperature of the ice sample was measured using a silicon diode sensor connected to the sample holder and placed directly above the ice sample. A Bayard-Alpert gauge positioned about 23 cm below the plane where deposition occurs was used to measure the pressure in the main chamber of ISAC. Laser interference was monitored continuously, resulting in an interference pattern during ice deposition and a new interference pattern during TPD. The ice column density was calculated using IR spectroscopy from the areas calculated by integration of the absorption bands. This is covered in more detail in the following section.
   
\subsection{Estimation of the ice mixture ratio}
\label{mixture ratio}
The integration of the infrared absorption band yields the column density $N$ of the ice layer accreted on the cold substrate in molecules cm$^{-2}$ with the following formula:
\begin{equation}
    N=\int_{band} \frac{\tau_{\nu}d\nu}{A},
\end{equation}
with $A$ the band strength in cm molecule$^{-1}$, $\tau_{\nu}$  the optical depth of the band and d$\nu$ the wavenumber differential in cm$^{-1}$. The integrated absorbance is equal to 0.43 $\times$ $\tau$, where $\tau$ is the integrated optical depth of the band. The ice mixture ratio is calculated by dividing the column density of the carbon dioxide anti-symmetric stretching band to the water stretching band.
The IR band strength of the CO$_2$ str. feature in the H$_2$O:CO$_2$ = 24:1 ice mixture is about 94\% with respect to the pure CO$_2$ ice (from Table 3 of \cite{Gerakines1995}), and for higher CO$_2$ concentrations this difference is expected to be smaller. We therefore used the band strengths of pure ices at 10 K: A(H$_2$O)=2.0$\times$10$^{-16}$ cm molecule$^{-1}$ (\cite{hagen}) and A(CO$_2$)=7.6$\times$10$^{-17}$ cm molecule$^{-1}$ (\cite{bouilloud}).
To compute the integrated intensities of the H$_2$O stretching peaks, the measured peak positions and the integration limits in cm$^{-1}$($\mu$m) are taken between 3000 (3.33) and 3600 (2.78). The same method is applied to the CO$_2$ anti-symmetric stretching mode in the 2330 (4.29) to 2360 (4.24) spectral range. A table summarizing the experimental parameters can be found in \autoref{table:1}, where $p_{dep}$ is the total pressure in the ISAC chamber during deposition of the ice, $T_{dep}$ is the temperature at which the ice is deposited, $d_{ice}$ is the ice thickness, and $dT/dt$ is the heating rate in Kelvin per minute. 



\begin{table}[h!]
\centering
\caption{Experimental parameters}             
\label{table:1} 
\begin{tabular}{c c c c c}
\hline
\hline
Composition & \multicolumn{1}{c}{\begin{tabular}[c]{@{}c@{}}$p_{dep}$\\ (mbar)\end{tabular}} & \multicolumn{1}{c}{\begin{tabular}[c]{@{}c@{}}$T_{dep}$\\ (K)\end{tabular}} & \multicolumn{1}{c}{\begin{tabular}[c]{@{}c@{}}$d_{ice}$\\ ($\mu$m)\end{tabular}} & \multicolumn{1}{c}{\begin{tabular}[c]{@{}c@{}}$\frac{dT}{dt}$\\ (K/min)\end{tabular}} \\ \hline
Pure CO$_2$         & 2$\times$10$^{-7}$                                                     & 10                                                 & 0.62                                                   & 0.5                                                     \\
H$_2$O:CO$_2$=1:0.04 & 2$\times$10$^{-7}$                                                      & 10                                                  & 0.54                                                   & 0.2                                                       \\
H$_2$O:CO$_2$=1:0.25 & 2$\times$10$^{-7}$                                                      & 10                                                  & 0.67                                                  & 0.2                                                     \\
\hline                                                    
\end{tabular}
\end{table}


\section{Pure CO$_2$ ice}
\label{ch3}
Pure CO$_2$ ice was deposited at 10 K and warmed up until 100 K. The study of the CO$_2$ ice stretching band, commonly known as $\nu_3$, as well as the TPD are reported in Sections \ref{3.1} and \ref{3.2}, respectively.

\subsection{$\nu_3$ stretching band shape and position}
\label{3.1}
\begin{figure}[]
    \centering
    \includegraphics[width=\hsize]{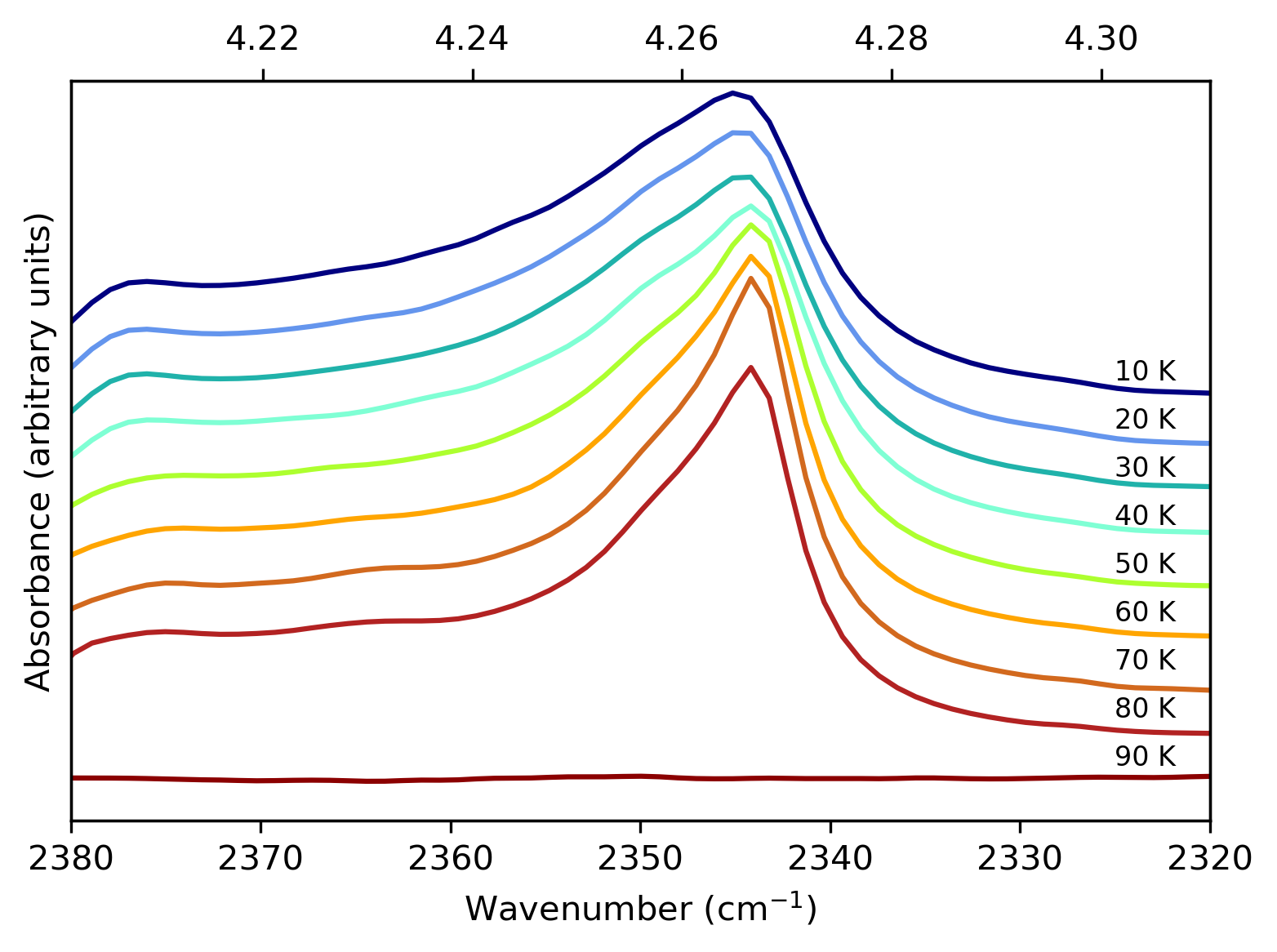}   
    \caption{Spectra over the 2380-2320 cm$^{-1}$ (4.20-4.31 $\mu$m) range of a pure CO$_2$ ice deposited at 10 K and 2$\times$10$^{-7}$ mbar, later warmed up to 90 K. All spectra are measured at 2 cm$^{-1}$ resolution, and at temperatures indicated in each graph. Spectra are shifted vertically for clarity.}
    \label{fig:pure CO2}
\end{figure}

\begin{figure}[]
    \centering
    \includegraphics[width=\hsize]{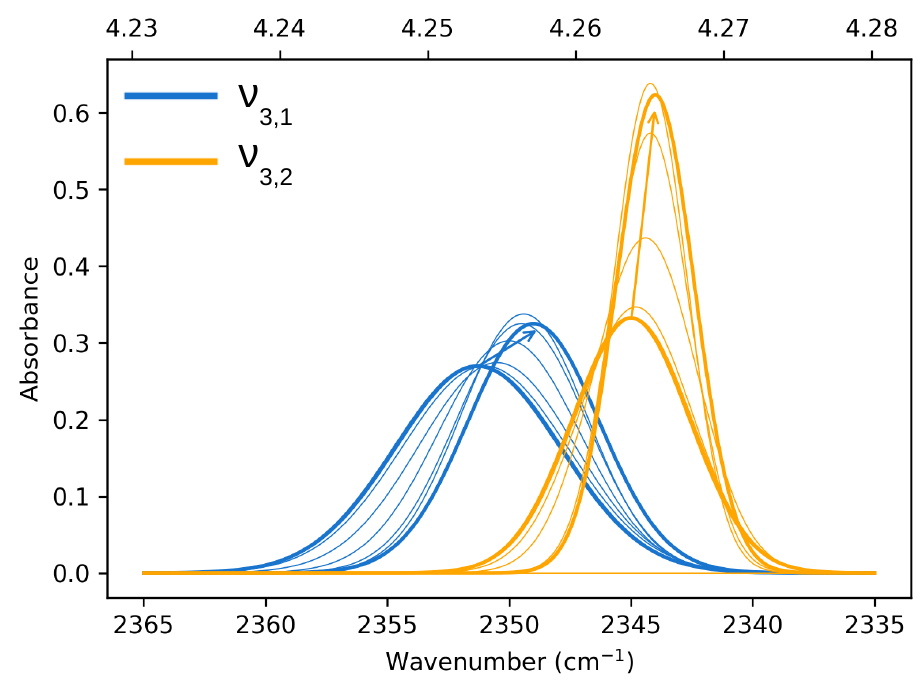}   
    \caption{Evolution of both Gaussian distributions during warm-up from 10 K to 80 K for a pure CO$_2$ ice. The arrows indicate the direction of evolution. Thicker lines indicate the initial and final states.}
    \label{fig:pure CO2_gaussians}
\end{figure}

Figure \ref{fig:pure CO2} shows the CO$_2$ ($\nu_3$) asymmetric stretching fundamental for $^{12}$CO$_2$ when deposited at 10 K, at a pressure of 2$\times$10$^{-7}$ mbar and warmed-up with a rate of 0.5 K/min up to 90 K. All the FTIR spectra have a resolution of 2 cm$^{-1}$ and are shifted vertically for clarity. The $^{12}$CO$_2$ asymmetric stretching fundamental ($\nu_3$) is located at $\sim$2345 cm$^{-1}$ (4.26 µm), and is redshifted from the gas-phase value located at 2348 cm$^{-1}$ (4.26 $\mu$m) due to interactions with the surrounding matrix environment (e.g. \cite{isokoski}). The profile is asymmetric with a prominent blue shoulder around 2350 cm$^{-1}$ (4.25 $\mu$m). When the CO$_2$ ice is warmed up, the main peak at 2345 cm$^{-1}$ (4.26 $\mu$m) shifts to lower wavenumbers. In addition, the peak intensity increases with increasing temperature as a result of the decrease in bandwidth. Figure \ref{fig:pure CO2_gaussians} illustrates the effect of temperature on the band's shape and position. 

In general, each spectrum can be deconvolved into two Gaussian distributions, which we refer to as $\tilde{\nu}_{3,1}$ at 2351 cm$^{-1}$ (4.25 $\mu$m) and $\tilde{\nu}_{3,2}$ at 2345 cm$^{-1}$ (4.26 $\mu$m), both for 10 K (see \autoref{tab:v}). The figure clearly shows the redshift of both Gaussians during warm-up as well as the decrease in FWHM for $\tilde{\nu}_{3,2}$ (2345 cm$^{-1}$, 4.26 $\mu$m), going from 5.7 cm$^{-1}$ at 10 K to 3.8 cm$^{-1}$ at 80 K.  

\begin{table*}[h!]
\caption{Band positions ($\tilde{\nu}$) for all three fitted Gaussians as a function of temperature. This is done for a pure CO$_2$ ice, a low CO$_2$ concentration and a high CO$_2$ concentration.}
\label{tab:v}
\centering
\begin{tabular}{cccccccccc}
\hline
\hline
\noalign{\smallskip}
      & \multicolumn{3}{c}{Pure CO$_2$} & \multicolumn{3}{c}{Low CO$_2$} & \multicolumn{3}{c}{High CO$_2$}          \\
T (K) & \begin{tabular}[c]{@{}c@{}}$\tilde{\nu}_{3,1}$\\ 
(cm$^{-1}$)\end{tabular} & \begin{tabular}[c]{@{}c@{}}$\tilde{\nu}_{3,2}$\\
(cm$^{-1}$)\end{tabular} & \begin{tabular}[c]{@{}c@{}}$\tilde{\nu}_{3,3}$\\ 
(cm$^{-1}$)\end{tabular} & \begin{tabular}[c]{@{}c@{}}$\tilde{\nu}_{3,1}$\\ 
(cm$^{-1}$)\end{tabular} & \begin{tabular}[c]{@{}c@{}}$\tilde{\nu}_{3,2}$\\ 
(cm$^{-1}$)\end{tabular} & \begin{tabular}[c]{@{}c@{}}$\tilde{\nu}_{3,3}$\\ 
(cm$^{-1}$)\end{tabular} & \begin{tabular}[c]{@{}c@{}}$\tilde{\nu}_{3,1}$\\ 
(cm$^{-1}$)\end{tabular} & \begin{tabular}[c]{@{}c@{}}$\tilde{\nu}_{3,2}$\\ 
(cm$^{-1}$)\end{tabular} & \begin{tabular}[c]{@{}c@{}}$\tilde{\nu}_{3,3}$\\ 
(cm$^{-1}$)\end{tabular} \\ 
\cline{1-10} 
\noalign{\smallskip}
10    & 2351.3 & 2345.0 & \multicolumn{1}{c|}{-}  & 2350.5  & 2345.0 & \multicolumn{1}{c|}{2340.8} &2353.5 &2344.3 &2337.3  \\
30    & 2351.0 & 2344.8 & \multicolumn{1}{c|}{-}  & 2350.2  & 2344.8 & \multicolumn{1}{c|}{2340.8} &2353.7 &2344.0 &2337.0  \\
50    & 2350.0 & 2344.3 & \multicolumn{1}{c|}{-}  & 2349.8  & 2344.5 & \multicolumn{1}{c|}{2340.7} &2353.8 &2344.0 &2337.1  \\
70    & 2349.4 & 2344.2 & \multicolumn{1}{c|}{-}  & 2349.8  & 2344.2 & \multicolumn{1}{c|}{2340.6} &2353.7 &2344.1 &2337.2  \\
80    & 2349.0 & 2344.0 & \multicolumn{1}{c|}{-}  & 2347.5  & -      & \multicolumn{1}{c|}{2340.7} &2351.5 &2344.1 &2339.0  \\
90    & -      & -      & \multicolumn{1}{c|}{-}  & 2347.0  & -      & \multicolumn{1}{c|}{2340.7} &2348.0 &-      &2339.7  \\
100   & -      & -      & \multicolumn{1}{c|}{-}  & 2347.0  & -      & \multicolumn{1}{c|}{2340.8} &2348.3 &-      &2340.0  \\
110   & -      & -      & \multicolumn{1}{c|}{-}  & 2347.0  & -      & \multicolumn{1}{c|}{2340.7} &2347.8 &-      &2339.9   \\
130   & -      & -      & \multicolumn{1}{c|}{-}  & 2346.9  & -      & \multicolumn{1}{c|}{2340.8} &2347.8 &-      &2339.8  \\
140   & -      & -      & \multicolumn{1}{c|}{-}  & 2347.0  & -      & \multicolumn{1}{c|}{2340.8} &2347.8 &-      &2339.9  \\
150   & -      & -      & \multicolumn{1}{c|}{-}  & 2346.8  & -      & \multicolumn{1}{c|}{2340.7} &2349.0 &-      &2340.3  \\
160   & -      & -      & \multicolumn{1}{c|}{-}  & 2348.0  & -      & \multicolumn{1}{c|}{2340.6} &2349.3 &-      &2340.5  \\
170   & -      & -      & \multicolumn{1}{c|}{-}  & 2348.0  & -      & \multicolumn{1}{c|}{-}      &2349.1 &-      &-                                                    
\end{tabular}
\end{table*}

\subsection{Thermal desorption of CO$_2$}
\label{3.2}
\begin{figure}[h!]
    \centering
    \includegraphics[width=\hsize]{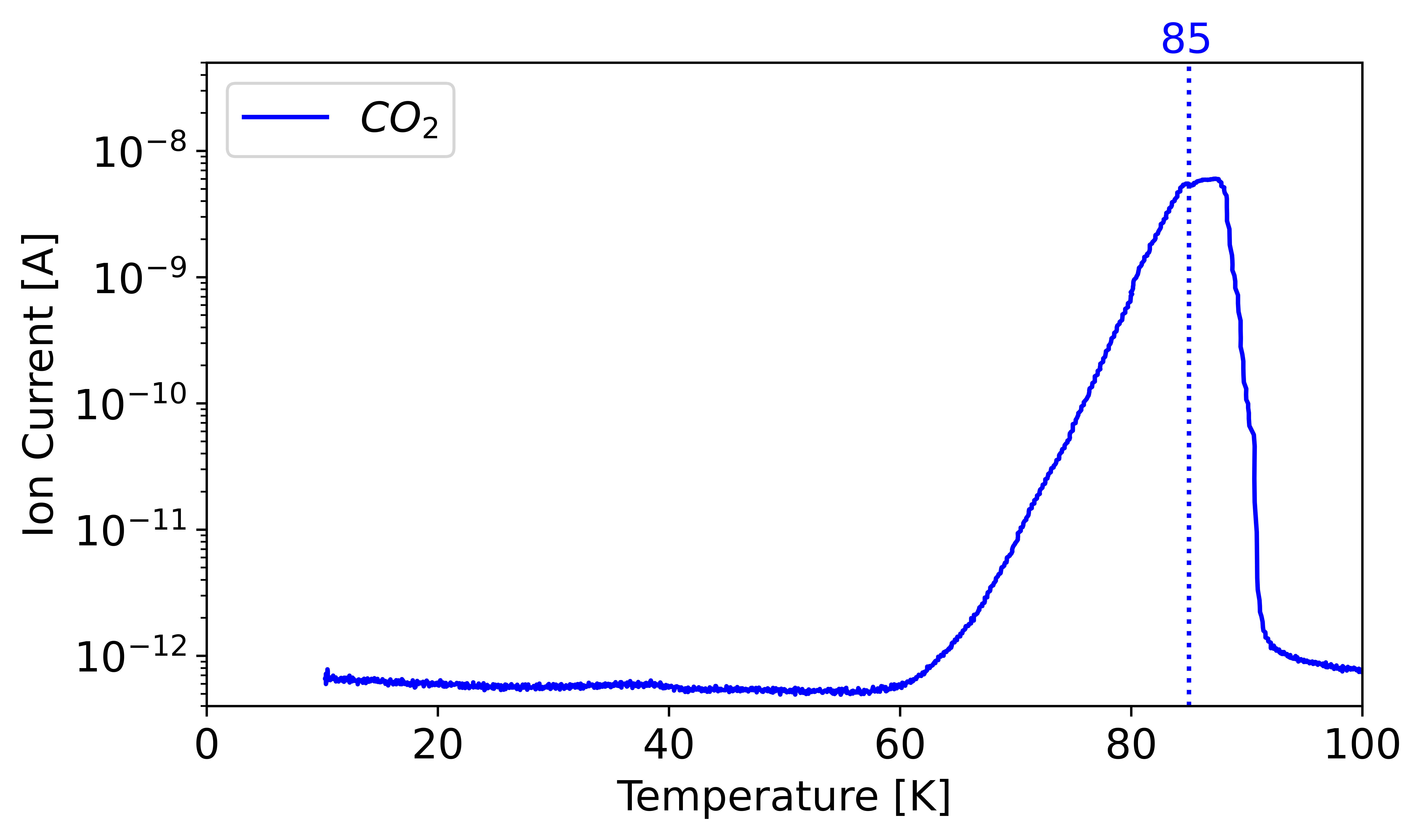}
    \caption{TPD curve of pure CO$_2$ ice layer deposited at 10 K and heated at 0.5 K/min. The ion current (A) is plotted on a logarithmic scale for a better appreciation of the curve profile and roughly corresponds to partial pressure in mbar.}
    \label{fig:QMS pure CO2}
\end{figure}

\begin{figure}[h!]
    \centering
    \includegraphics[width=\hsize]{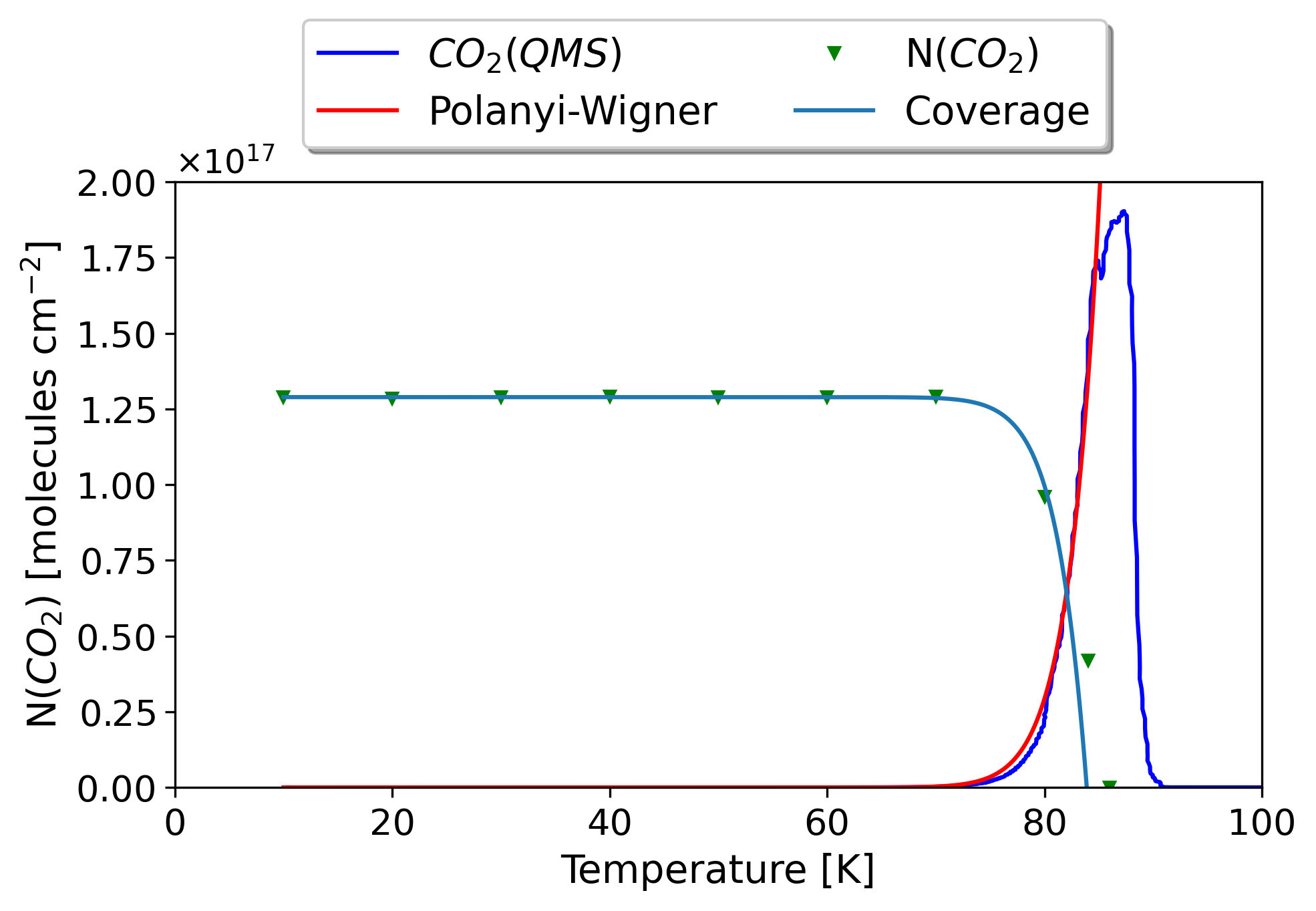}
    \caption{Column density of CO$_2$ ice (triangles) calculated from IR measurements matches the calculated coverage (light blue line). TPD curve of CO$_2$ from QMS measurements (dark blue) agrees well with the fit corresponding to the Polanyi-Wigner equation (red line).}
    \label{fig:PW Pure co2}
\end{figure}

A TPD curve of a pure CO$_2$ ice layer is shown in \autoref{fig:QMS pure CO2}. The QMS data corresponding to the molecular mass of CO$_2$ is plotted as the ion current in Ampere versus increasing temperature. This figure shows one desorption peak at 85 K. It is at this temperature that CO$_2$ thermally desorbs. The desorption rate, expressed in molecules cm$^{-2}$ s$^{-1}$, of CO$_2$ ice is described by the Polanyi-Wigner equation:
\begin{equation}
\label{PW}
    \frac{dN_g(CO_2)}{dt}=\nu_i[N_s(CO_2)]^i\text{exp}\left(-\frac{E_d(CO_2)}{T}\right),
\end{equation}
where $N_g(CO_2)$ is the column density of CO$_2$ molecules that desorb from the ice surface (cm$^{-2}$), $\nu_i$ is the frequency factor (molecules$^{1-i}$ cm$^{-2(1-i)}s^{-1})$ for desorption order, $i$, $N_s(CO_2)$ is the column density of CO$_2$ molecules on the surface at time $t$, $E_d(CO_2)$ is the binding energy expressed in K, and $T$ the surface temperature in K. The TPD data can be fitted using \autoref{PW} and the relation:
\begin{equation}
    \frac{dN_g(CO_2)}{dt}=\frac{dT}{dt}\frac{dN}{dT},
\end{equation}
where $\frac{dT}{dt}$ is the heating rate, 0.5 K min$^{-1}$ in the reported experiments. 
Figure \ref{fig:PW Pure co2} shows the QMS data, column densities during warm-up, Polanyi-Wigner equation, and calculated zero-order coverage. The coverage is calculated using 1-$\frac{dN_g(CO_2)}{dt}$. The TPD data were fitted using the parameter values $\nu_0$ = 0.4 $\times$ 10$^{31}$ molecules cm$^{-2}$ s$^{-1}$ and $E_d$ = 2555.78 K for the desorption of CO$_2$ ice. We find that the coverage values determined by transmittance FTIR (green triangles) and the coverage curve computed by fitting the TPD curve recorded by QMS (light blue line) are in good agreement. There are only two remaining triangle datapoints that are poorly fitted. The issue is that a zero-order fit by definition fails when there are fewer than 4 $\times$ 10$^{16}$ molecules per square centimeter of coverage, as demonstrated by \autoref{fig:PW Pure co2}.

\section{Low CO$_2$ concentration in water ice}
\label{ch4}
A low CO$_2$ content in the ice mixtures, up to 5$\%$, is considered in this section. The important parameters of the presented experiments can be found in \autoref{table:1}. Figure \ref{fig:low c, low p} depicts the spectra for a H$_2$O:CO$_2$=1:0.04 ratio deposited at 10 K. The experiment is performed with a deposition pressure of 2$\times$10$^{-7}$ mbar as reported in \autoref{table:1}. At 10 K, the $^{12}$CO$_2$ asymmetric stretching fundamental ($\nu_3$) can be deconvolved into three Gaussians in these experiments, $\tilde{\nu}_{3,1}$ around 2351 cm$^{-1}$ (4.25 $\mu$m), $\tilde{\nu}_{3,2}$ at 2345 cm$^{-1}$ (4.26 $\mu$m), and $\tilde{\nu}_{3,3}$ at 2341 cm$^{-1}$ (4.27 $\mu$m).
The integrated absorbance area as a function of temperature is shown in the bottom panel of \autoref{fig:QMS, 4 percent}. 
Blue dots represent $\tilde{\nu}_{3,1}$ (2351 cm$^{-1}$, 4.25 $\mu$m) starting at 10 K, which undergoes a red shift at elevated temperatures. Yellow triangles represent $\tilde{\nu}_{3,2}$ (2345 cm$^{-1}$, 4.26 $\mu$m), which is red shifted by 1 cm$^{-1}$ at 80 K right before thermal desorption.
Green crosses representing $\tilde{\nu}_{3,3}$ (2341 cm$^{-1}$, 4.27 $\mu$m) indicate no observable changes with increasing temperature. 

\begin{figure}[h!]
    \centering
    \includegraphics[width=\hsize]{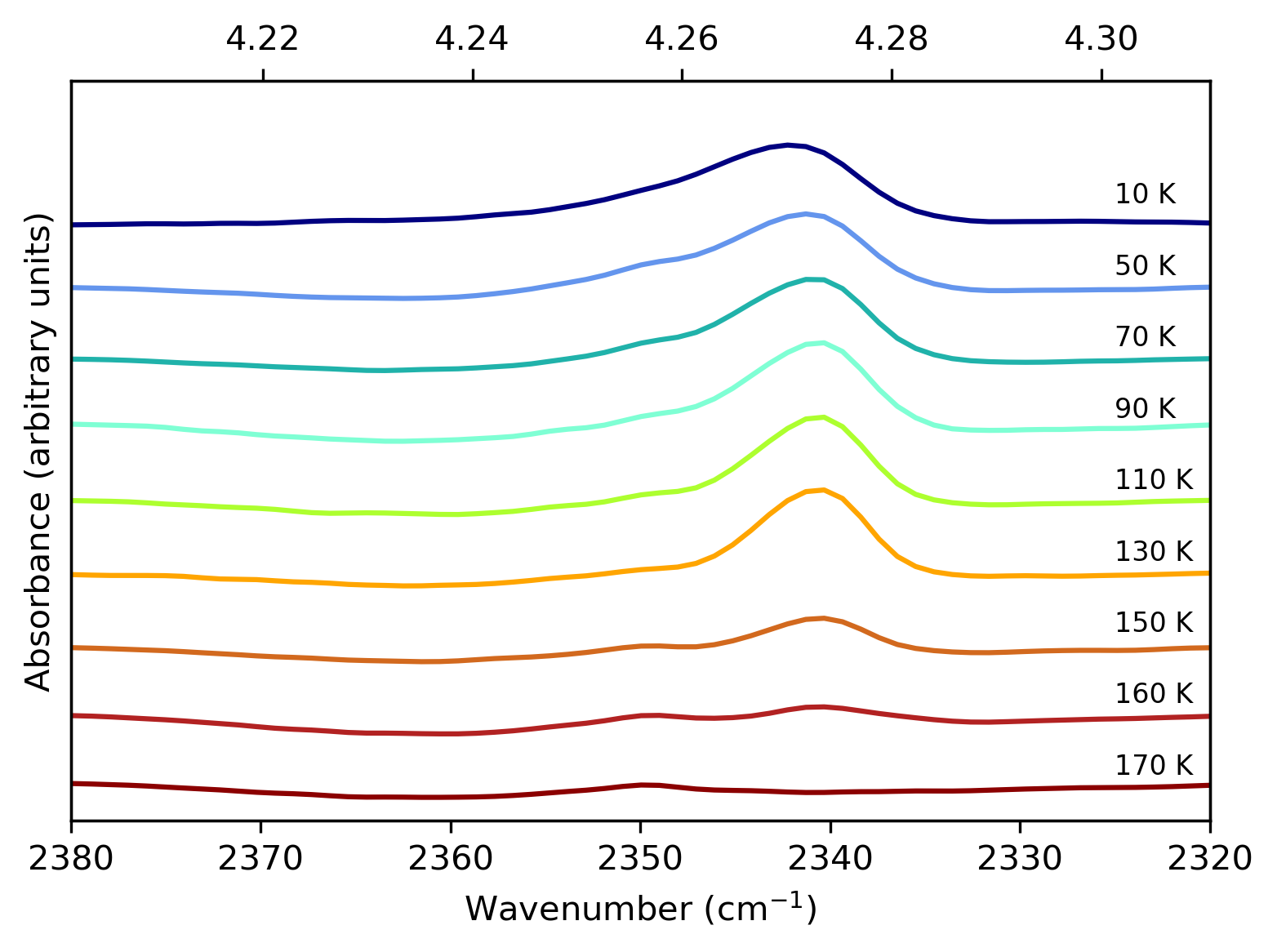}
    \caption{Spectra over the 2380-2320 cm$^{-1}$ 
    (4.20-4.31 $\mu$m)
    range of the H$_2$O:CO$_2$=1:0.04 ice mixture deposited at 10 K and 2$\times$10$^{-7}$ mbar. All spectra are measured with 2 cm$^{-1}$ resolution at the temperatures indicated in each graph. Spectra are shifted vertically for clarity.}
    \label{fig:low c, low p}
\end{figure}

As mentioned before, pure CO$_2$ ice thermally desorbs around 85 K. For the ice mixture H$_2$O:CO$_2$=1:0.04, the top panel of \autoref{fig:QMS, 4 percent} shows the TPD curves from 10 K to 180 K, with three desorption peaks for carbon dioxide: one around 80 K, another one around 146 K, and a final one at 162 K, while only one peak appears for water, also at 162 K. 
There is a smaller CO$_2$ desorption peak at 150 K which is not reproducible in repeated experiments and could be caused by a minor rearrangement of the water structure.
The following explanation is given for the three main desorption peaks of carbon dioxide. Firstly, pure carbon dioxide ice thermally desorbs around 80 K, meaning that only CO$_2$ molecules bonded with other CO$_2$ molecules are desorbing.
In second place, the peak at 140 K is caused by the transformation from cubic to hexagonal water ice \citep{domenech}. During this process, carbon dioxide is pushed out of the water matrix. This peak is known as volcano desorption.
The third peak between 160 and 170 K is the co-desorption of all remaining carbon dioxide along with water.

It is worth noting that carbon dioxide constantly desorbs between 80 and 146 K, as evidenced by the QMS signal being greater after the initial peak than before (see blue line in top panel of \autoref{fig:QMS, 4 percent}). This continuous desorption is caused by amorphous water ice rearrangement, which produces a change in the size and form of the porous structure (pores coalesce) when heated \citep{cazaux}. Carbon dioxide molecules may diffuse through the pores and sublimate continuously between 80 and 146 K due to the increasing ice temperature and favored by this rearrangement. The argument of diffusing carbon dioxide molecules can be verified by looking at the integrated absorbance areas of the three Gaussian peaks in the bottom panel of \autoref{fig:QMS, 4 percent}. 
The data points (triangles) represent $\tilde{\nu}_{3,2}$ (2345 cm$^{-1}$, 4.26 $\mu$m) and confirm that pure CO$_2$ desorbs thermally at 80 K. 
During this desorption of CO$_2$ molecules starting at 75 K, the integrated area of $\tilde{\nu}_{3,3}$ (2341 cm$^{-1}$, 4.27 $\mu$m) is increasing significantly, pointing towards the diffusion of the carbon dioxide molecules through the water ice structure. 
Once temperatures of 140 K are reached, the integrated areas of both $\tilde{\nu}_{3,1}$ (2351 cm$^{-1}$, 4.25 $\mu$m) and $\tilde{\nu}_{3,3}$ (2341 cm$^{-1}$, 4.27 $\mu$m) bands are continuously decreasing during annealing, confirming the amorphous water ice rearrangements. The cubic to hexagonal water-ice phase transition that occurs at 146 K further accelerates the desorption of CO$_2$ molecules. Finally, any remaining CO$_2$ will co-desorb with water around 162 K.

\begin{figure}[h!]
    \centering
    \includegraphics[width=\hsize]{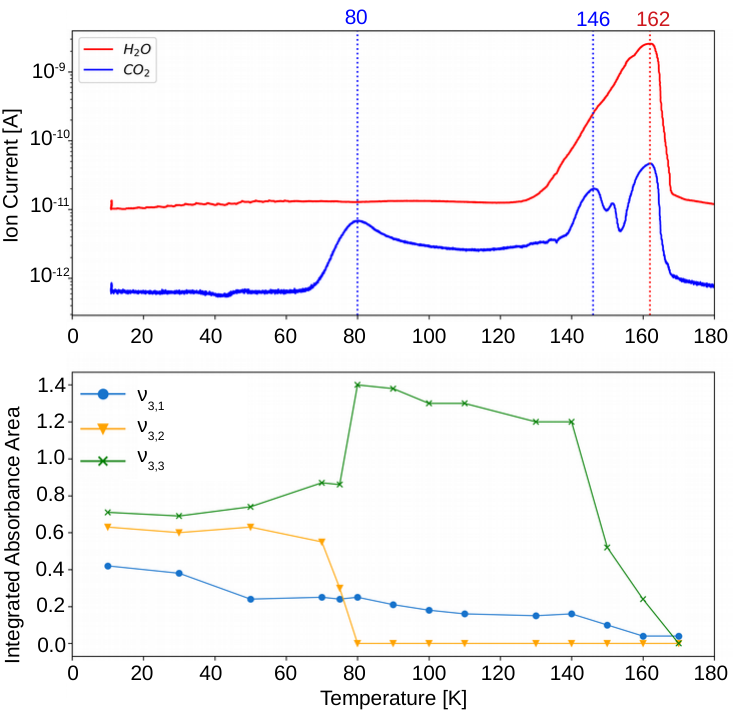}
    \caption{TPD curves of CO$_2$ and H$_2$O for a H$_2$O:CO$_2$=1:0.04 ice heated at 0.2 K/min (top).  Integrated absorbance areas of the three Gaussian distributions as a function of temperature for the same experiment (bottom).}
    \label{fig:QMS, 4 percent}
\end{figure}

\section{High CO$_2$ concentration in water ice}
\label{ch5}
In this section, a high CO$_2$ content in the ice mixtures, above 20$\%$, is considered. The parameters of the conducted experiments can be found in \autoref{table:1}.
Figure \ref{fig:low c, high pressure} shows the spectra for H$_2$O:CO$_2$=1:0.25 ice deposited at 10 K and 2$\times$10$^{-7}$ mbar. The broadening of the band is a consequence of the increased CO$_2$ concentration, which causes particle aggregation and thus creates additional trapping sites between CO$_2$ and H$_2$O (\cite{ehrenfreund}). As with the low CO$_2$ concentration, the spectra can be deconvolved into three Gaussian distributions, the band positions at 10 K are $\tilde{\nu}_{3,1}$ = 2354 cm$^{-1}$ (4.25 $\mu$m), $\tilde{\nu}_{3,2}$ = 2344 cm$^{-1}$ (4.27 $\mu$m), and $\tilde{\nu}_{3,3}$ = 2336 cm$^{-1}$ (4.28 $\mu$m), at 10 K. Due to the broadening of the CO$_2$ band, the shoulder on the left is blueshifted by 2 cm$^{-1}$, compared to the low CO$_2$ concentration spectra at 10 K. In addition, the shoulder on the right is redshifted by 4 cm$^{-1}$.   

\begin{figure}[h!]
    \centering
    \includegraphics[width=\hsize]{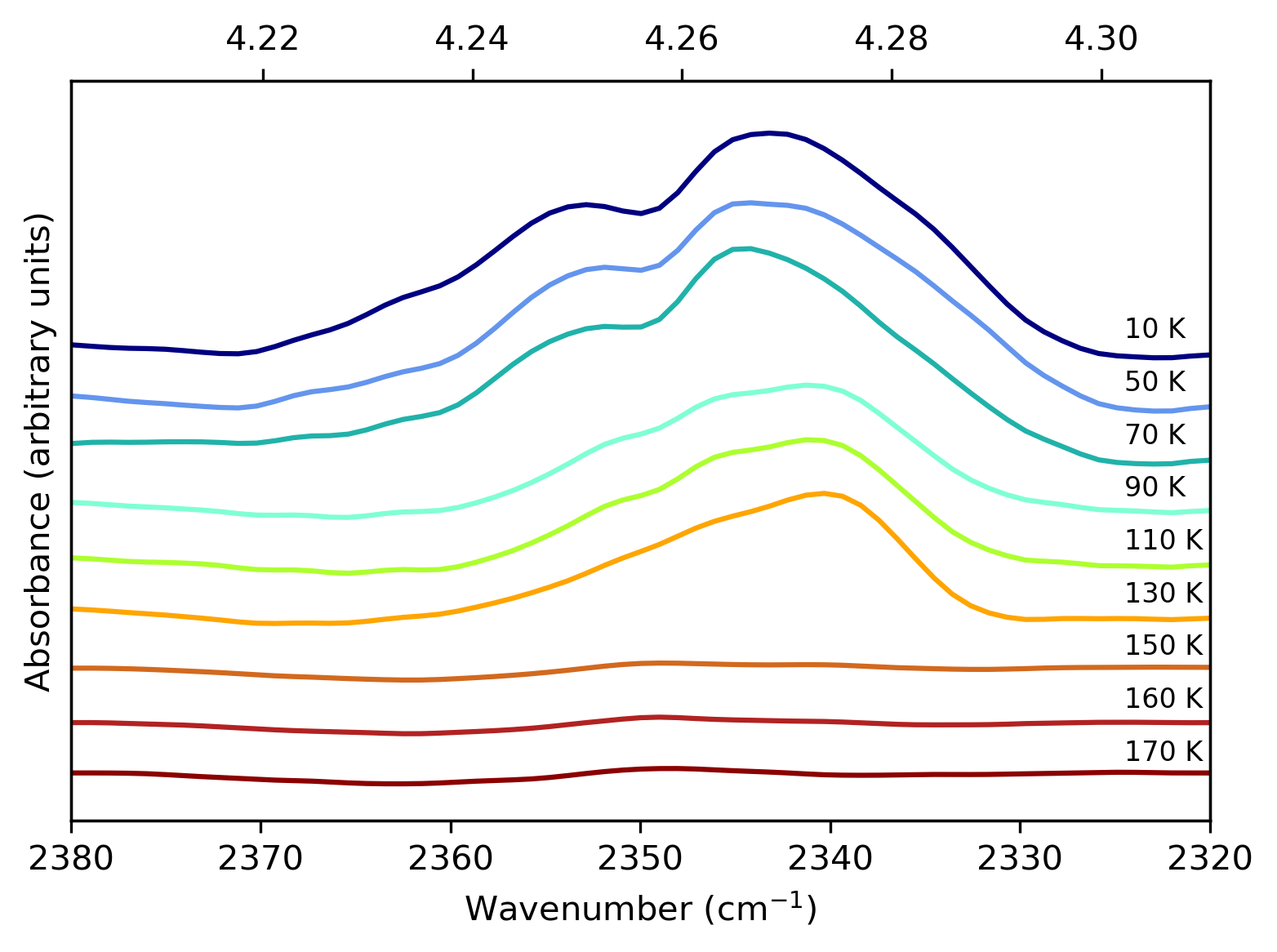}
    \caption{Spectra over the 2380-2320 cm$^{-1}$ (4.20-4.31 $\mu$m) range of the H$_2$O:CO$_2$=1:0.25 ice mixture deposited at 10 K and 2$\times$10$^{-7}$ mbar. All spectra are measured with 2 cm$^{-1}$ resolution at the temperatures indicated in each graph. Spectra are shifted vertically for clarity.}
    \label{fig:low c, high pressure}
\end{figure}
\noindent The TPD curves and integrated absorbance area are shown in \autoref{fig:high c, low pressure}. Again, three desorption peaks for carbon dioxide can be identified, namely at 82 K, 146 K, and 161 K. The same explanation as for a low CO$_2$ concentration is valid for this experiment. The small bump in the TPD curve around 30 K is not reproducible in repeated experiments and might be due to a small amount of air in the gas line. 
The TPD curve of carbon dioxide shows the main desorption at 82 K, corresponding to pure CO$_2$ desorption in these experiments, as is confirmed by the drastic drop of the integrated absorbance area of $\tilde{\nu}_{3,2}$ at that temperature.
The bottom panel of \autoref{fig:high c, low pressure} clearly shows, as in the case of low CO$_2$ concentration, the diffusion through the water ice structure after the first desorption peak, indicated by the increase in integrated area for $\tilde{\nu}_{3,3}$ 
The desorption of CO$_2$ is enhanced by the phase transition from cubic to hexagonal water ice at 146 K and any remaining CO$_2$ co-desorbs with water at 161 K.

\begin{figure}[h!]
    \centering
    \includegraphics[width=\hsize]{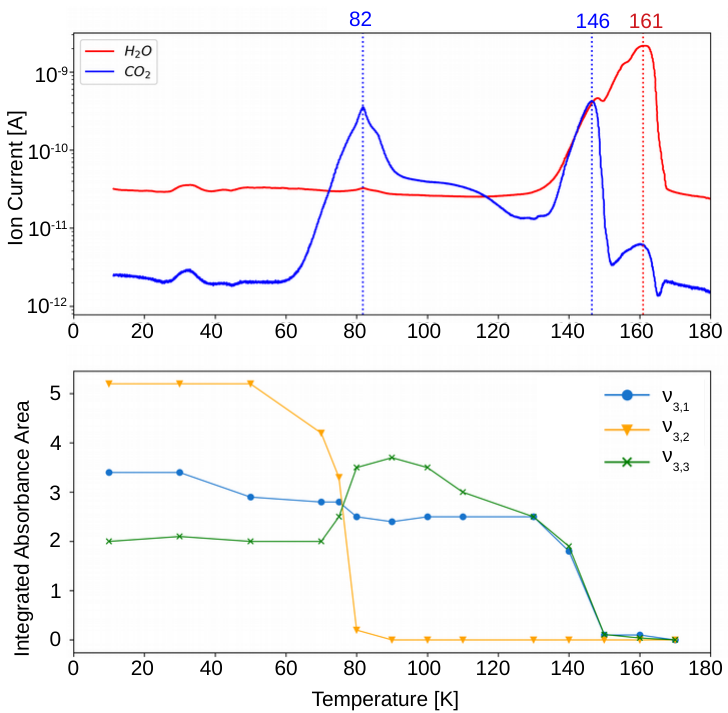}
    \caption{TPD curves of CO$_2$ and H$_2$O for a H$_2$O:CO$_2$=1:0.25 ice layer heated at 0.2 K/min (top). Integrated absorbance areas of the three Gaussian distributions as a function of temperature for a H$_2$O:CO$_2$=1:0.25 ice (bottom).}
    \label{fig:high c, low pressure}
\end{figure}
\section{Infrared band assignments}
\label{sec:discussion}
\autoref{tab:v} provides an overview of the positions of the fitted Gaussians of the asymmetric stretching bands, denoted as $\tilde{\nu}_{3,1}$, $\tilde{\nu}_{3,2}$, and $\tilde{\nu}_{3,3}$. This was done for a pure CO$_2$ ice, low or high CO$_2$ concentration in water ice. These measurements span from 10 K (the deposition temperature) to 170 K, which corresponds to the desorption temperature of water in these experiments.\\
For pure CO$_2$ ice, the positions of $\tilde{\nu}_{3,1}$ and $\tilde{\nu}_{3,2}$, respectively 2351 cm$^{-1}$ (4.25 $\mu$m) and 2345 cm$^{-1}$ (4.26 $\mu$m) at 10 K, are affected by temperature, shifting towards the red as the temperature increases (as also demonstrated in \autoref{fig:pure CO2_gaussians}). A similar behaviour is observed both when low and high CO$_2$ concentrations are mixed with water. We note that the $\tilde{\nu}_{3,2}$ (at 10 K it falls at 2345 cm$^{-1}$ or 4.26 $\mu$m) shift for high CO$_2$ concentration is much smaller than for the low concentration. At 10 K, a result of $\tilde{\nu}_{3,1}$ (2351 cm$^{-1}$, 4.25 $\mu$m) for a low CO$_2$ concentration closely aligns with the position in pure CO$_2$ ice, whereas a high concentration results in a blueshift of over 2 cm$^{-1}$. In addition, $\tilde{\nu}_{3,2}$ at high concentration (2344 cm$^{-1}$, 4.27 $\mu$m) is redshifted by 0.7 cm$^{-1}$ compared to the low concentration and pure CO$_2$ ice. The $\tilde{\nu}_{3,3}$ appears only when mixed with water. The peak position of $\tilde{\nu}_{3,3}$ (2337 cm$^{-1}$, 4.28 $\mu$m) in high CO$_2$ concentration experiences a redshift of nearly 4 cm$^{-1}$, compared to low concentration. These shifts are due to the band broadening (\cite{ehrenfreund}), and cause a shift of the Gaussian fits that is observed when the CO$_2$ concentration is increased in water ice.

\begin{figure}[h!]
    \centering
    \includegraphics[width=\hsize]{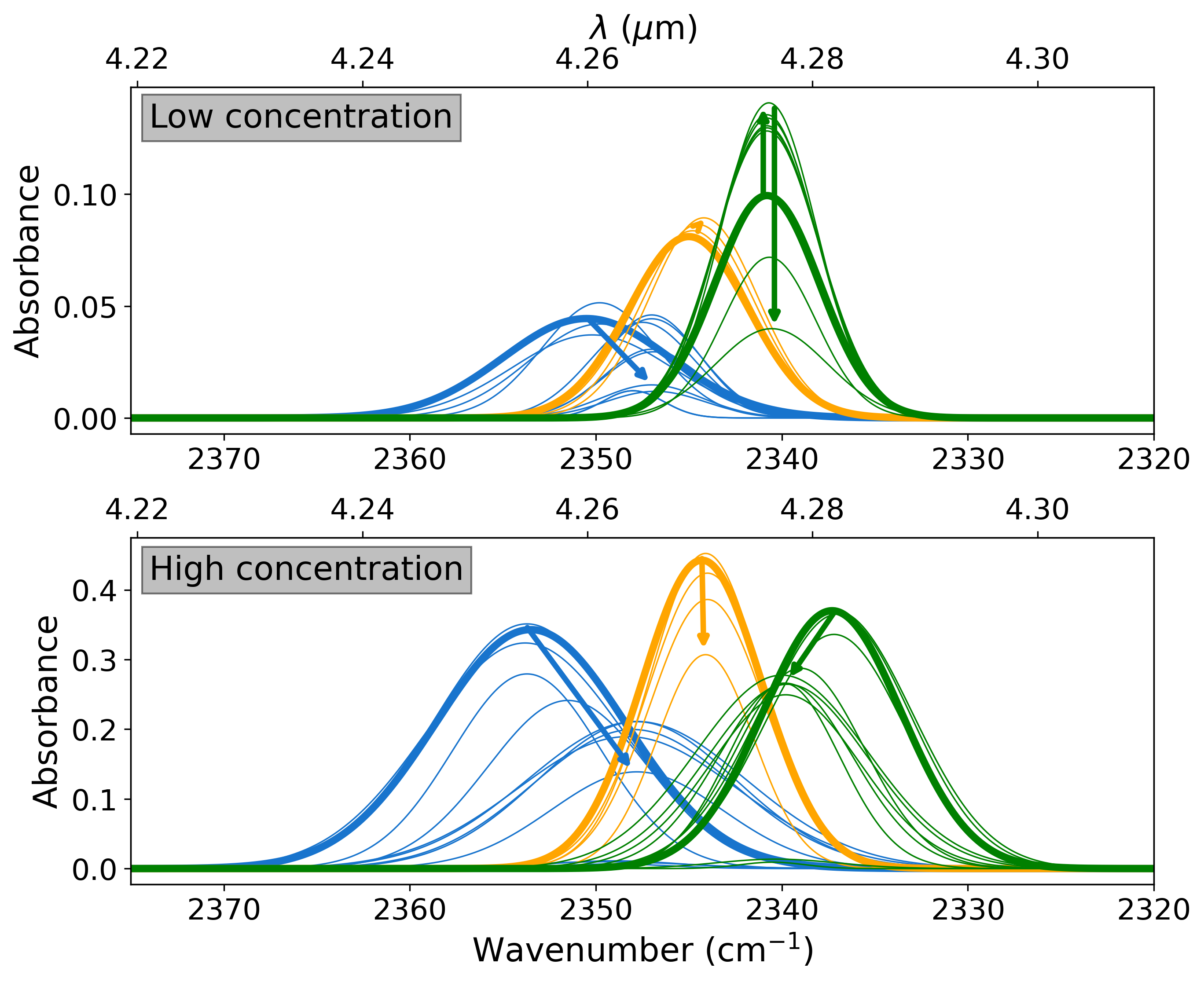}
    \caption{Three Gaussian distributions plotted from 10 K to 170 K, for a low and high CO$_2$ concentration in the ice. Blue represents $\tilde{\nu}_{3,1}$, orange $\tilde{\nu}_{3,2}$, and green $\tilde{\nu}_{3,3}$. The bold graphs indicate the beginning of the experiment at 10 K. The arrows indicate the direction of evolution.}
    \label{fig:summ gaussians}
\end{figure}

When mixed with water, the thermal evolution of the CO$_2$ $\tilde{\nu}_{3,1}$, $\tilde{\nu}_{3,2}$, and $\tilde{\nu}_{3,3}$ are reported in \autoref{fig:summ gaussians}. This figure depicts the three Gaussians from 10 K to 170 K. The top panel shows a low CO$_2$ concentration whereas the bottom panel corresponds to a high concentration. The Gaussians for 10 K are highlighted for clarity representing the beginning of the experiment. The 
arrows indicate the direction of evolution with increasing temperature, clearly illustrating how $\tilde{\nu}_{3,3}$ (2337 cm$^{-1}$, 4.28 $\mu$m) shifts to the blue only for a high CO$_2$ concentration. For a low concentration, the peak intensity of $\tilde{\nu}_{3,3}$ (2341 cm$^{-1}$ at 10 K, or 4.27 $\mu$m) first increases after 80 K, which becomes CO$_2$ embedded in the water ice matrix, as illustrated in the top panel of \autoref{fig:summ gaussians}. At 80 K, near the desorption temperature of pure CO$_2$, $\tilde{\nu}_{3,1}$ shifts to lower wavenumbers for both low and high CO$_2$ concentrations, their respective positions at this temperature are about 2348 cm$^{-1}$ (4.26 $\mu$m) and 2352 cm$^{-1}$ (4.25 $\mu$m). At 140 K, the integrated absorbance areas decrease due to the transition from cubic to hexagonal ice structure (\cite{domenech}). It is worth noting that in \autoref{tab:v}, for temperatures above 160 K, $\tilde{\nu}_{3,1}$ shifts back towards the blue, ultimately resulting in a single peak at around 2348 cm$^{-1}$ 
(4.259 $\mu$m) at 170 K for low and 2349.1 cm$^{-1}$ (4.257 $\mu$m) for high CO$_2$ concentration.
Furthermore, we explored the influence of deposition pressure on the band positions. To investigate this aspect, an experiment was conducted with a deposition pressure of 8$\times$10$^{-7}$ mbar and a 4$\%$ CO$_2$ concentration. Upon analyzing the band positions at 10 K, it was observed that the positions of $\tilde{\nu}_{3,2}$ (2345 cm$^{-1}$, 4.26 $\mu$m) and $\tilde{\nu}_{3,3}$ (2341 cm$^{-1}$, 4.27 $\mu$m) were consistent with the experiment conducted at a lower deposition pressure. 
The higher deposition pressure is known to induce changes in the ice morphology, rendering it more porous \citep{Mate2008,Bossa2015}. 
The DFT simulations detailed in \autoref{sec:DFT} show the effect porosity has on amorphous CO$_2$ ice, allowing the CO$_2$ dangling bond to occur. As the ice produced at higher pressure is more porous, it undergoes greater reorganization during the warm-up process. 
At 10 K, $\tilde{\nu}_{3,1}$ has a small $\sim$1 cm$^{-1}$ blue shift compared to the low deposition pressure. 
When the temperature reached 50 K, CO$_2$ became crystalline and the reorganization was complete. At this point, $\tilde{\nu}_{3,1}$, $\tilde{\nu}_{3,2}$, and $\tilde{\nu}_{3,3}$ were all peaking at approximately 2349.8 cm$^{-1}$ (4.256 $\mu$m), 2344.5 cm$^{-1}$ (4.265 $\mu$m), and 2340.6 cm$^{-1}$ (4.272 $\mu$m), respectively. These band positions are similar to the low deposition pressure experiment. The assignment of the different band components is discussed in the remainder of this section. As previously mentioned, when CO$_2$ is mixed with water, three distinct peaks emerge.

The $\tilde{\nu}_{3,2}$ band found at approximately 2345 cm$^{-1}$ (4.26 $\mu$m) can be attributed to bulk CO$_2$ ice. This is corroborated by the desorption temperature of pure CO$_2$, which falls in the range of 80-85 K for both low and high CO$_2$ concentrations. This desorption is illustrated in \autoref{fig:QMS pure CO2}, with band positions shown in \autoref{tab:v}. 
This band is also referred to as the CO$_2$-\textit{ext} where the interaction with water is quite weak as expected if CO$_2$ is superficially adsorbed, as proposed by \cite{galvez}.

The $\tilde{\nu}_{3,3}$ band arises when CO$_2$ is mixed with water and peaks around 2340 cm$^{-1}$ (4.27 $\mu$m). This band is the consequence of CO$_2$ molecules interacting with the water ice matrix. This peak is also referred to as CO$_2$-\textit{int} in previous work \citep{galvez,Mate2008}, where individual CO$_2$ molecules are trapped in the amorphous water ice and this interaction results in a slight weakening of the C-O bond, which produces a redshift on the IR spectrum \citep{sandford}. When increasing the temperature the integrated absorbance area of $\tilde{\nu}_{3,3}$ near 2340 cm$^{-1}$ (4.27 $\mu$m) diminishes as the water ice evolves to a hexagonal structure and the CO$_2$ molecules desorb from the water ice matrix.  

Here, $\tilde{\nu}_{3,1}$ was found at approximately 2351 cm$^{-1}$ (4.254 $\mu$m) for low CO$_2$ concentration and around 2353 cm$^{-1}$ (4.250 $\mu$m) for high concentration. 
This band is present in pure CO$_2$ ice, meaning that it is not caused by the interaction with water but is instead a consequence of the morphology of the ice. 
The band has been previously observed in ice growth experiments with high deposition rates \citep{Galvez2008}, although it was not observable at lower deposition rates \citep{Falk1987}. 
We propose that this band is caused by pores in the ice where the binding is particularly weak, as it occurs on the top surface of the ice, and the CO$_2$ molecules are dangling from the pore surface. This preliminary assignment is supported by a vibrational mode that is connected to porosity, namely, a degree of freedom to vibrate that is not allowed in a compact CO$_2$ ice structure; furthermore, the 2351 cm$^{-1}$ position relative to that of the fundamental C=O stretch in CO$_2$ ice is analogous to the relative position of the O-H dangling with respect to the O-H stretch in water ice \citep{matsuda}.This phenomenon is further supported by the discussion in \autoref{sec:DFT}.   Therefore, we assigned $\tilde{\nu}_{3,1}$ to the CO$_2$ dangling bonds. When CO$_2$ is mixed with water, this peak is also present; however, in this case, it remains above the 80 K temperature, which implies that following the desorption of bulk CO$_2$ ice, carbon dioxide molecules can remain trapped in the pores of the amorphous solid water.

A similar explanation for this band location, involving molecules bonded to -OH dangling bonds in the water ice, was suggested by \cite{matsuda} for CO molecules. In that case, two bands are assigned, namely CO molecules interacting with the OH dangling groups (CO-dangling OH) and CO molecules interacting with the oxygen atoms of the surface water molecules (CO-bonded OH). It was found that the area of the CO-dangling band increases when cycling to higher temperatures, meaning that CO molecules partially diffuse to CO-dangling OH groups on the surface of the amorphous water ice pores. 

The position of this band, blueshifted from the pure CO$_2$ stretching band, could be easily confused with a similar band observed in clathrate hydrates \citep{dartois2009}. In our case we can discard these cage-like structures as they are normally formed at much slower deposition rates and higher temperatures.

Other proposed explanations for this band location is that carbon dioxide is in a ``complexed'' form with water, as suggested by \cite{chaban}. A blue shift of 5 cm$^{-1}$ compared to the pure CO$_2$ ice is observed when bonded with one water molecule, and 10 cm$^{-1}$ with two. Furthermore, the symmetric O-H stretching band undergoes a redshift when complexed with two water molecules and the intensity increases significantly. In this work, for a 4$\%$ CO$_2$ concentration, there was no increase in intensity nor a redshift at 10 K for the symmetric O-H stretching band when compared to that of a pure H$_2$O ice, ruling out the possibility of a complexed CO$_2$. 

\section{Assignment of $\nu_{3.1}$ with DFT calculations}
\label{sec:DFT}
In order to better understand the influence of porosity in amorphous CO$_2$, simulations were carried out using density functional theory (DFT) \citep{DFT-HK,DFT-KS} using the CASTEP program \citep{CASTEP} for geometry optimization and vibrational spectrum prediction.

Starting with a CO$_2$ crystal, the amorphous ice models are created by applying temperature using molecular dynamics (MD) in the NPT ensemble until the structure melts. This was done with the Andersen barostat \citep{Andersen1980} at constant pressure and the Nose thermostat \citep{NHC} at rising temperatures. The porous model is then formed by removing CO$_2$ molecules from the amorphous model’s core. DFT was then used to geometrically optimize the systems using the generalized gradient approximation (GGA) and functionals by Perdew-Burke-Ernzerhof (PBE). Infrared spectra were simulated using density functional perturbation theory \citep{CASTEP_DFPT} based on these improved models. 

The calculated spectra are not intended to perfectly duplicate the measured wavenumbers for vibrational bands, but should be near within a margin of error. Most significantly, they are repeatable and can aid in understanding the influence of experimentally uncontrollable factors.

\begin{figure}[]
    \centering
    \includegraphics[width=\hsize]{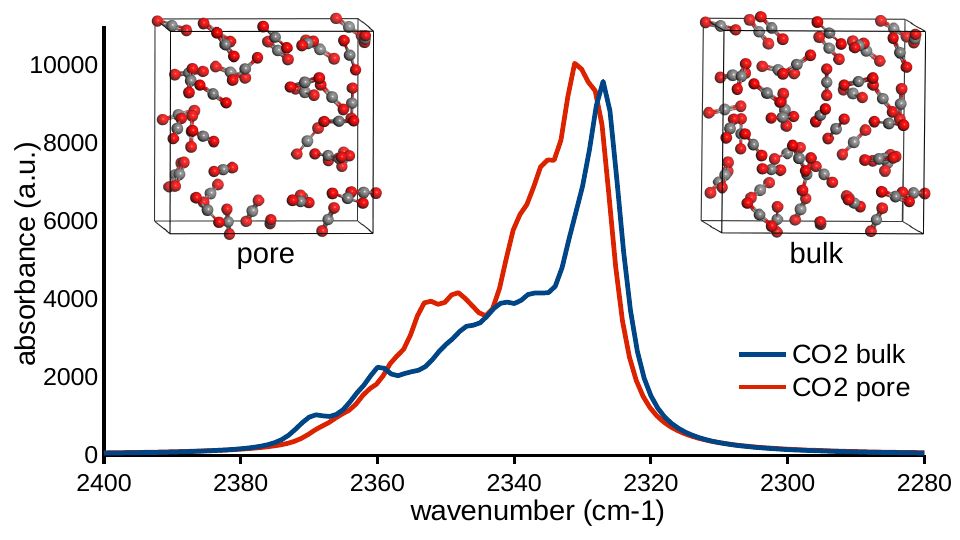}
    \caption{DFT calculated spectra for bulk amorphous CO$_2$ ice (blue) and a simulated pore in the same system ({orange}) as a function of wavenumber.}
    \label{fig:DFT}
\end{figure}

The simulated IR spectra of an amorphous CO$_2$ model (blue line) and a porous CO$_2$ model ({orange} line) are shown in \autoref{fig:DFT}. The morphology of the bulk ice model (right) and the porous model (left) are included for comparison. In the case of the bulk amorphous ice, we can see that the spectrum is very similar to the experimental case, with some expected offset for the wavenumber. In the porous model there is a clear blueshift of appproximately 4 cm$^{-1}$ for the main peak. Most importantly, a new secondary peak emerges at higher wavenumbers. This pattern is very similar to the spectra seen in laboratory ice, as well as the recent observations of this same CO$_2$ band in Ganymede. 
It must be stressed that the goal of this simulation is to simulate the effect of porosity on the peak location and shape. Therefore, finding the mechanism behind the shift rather than matching perfectly the band location with the same wavenumbers.

\section{Applications to icy moons} 
\label{ch7}
In this section, we use JWST ERS data from Ganymede \citep{bockelee} and Europa \citep{villanueva} to study the position and evolution of the CO$_2$ bands with varying latitude and longitude. 
{The data were reprocessed from the Mikulski Archive for Space Telescopes (MAST) as described in \cite{bockelee}.}
Our goal is to assign the CO$_2$ bands observed to the physical state of CO$_2$ ice. On Ganymede, the amount of CO$_2$ relative to water in mass is of the order of 1$\%$ \citep{bockelee}, which implies that our low concentration experiments are more representative of these icy moons' surfaces.

\subsection{Ganymede}
This section offers a discussion of the latitude and longitudinal spectra of Ganymede's leading hemisphere, as similar trends are observed on the trailing hemisphere.

\begin{figure}[h!]
    \centering
    \includegraphics[width=\hsize]{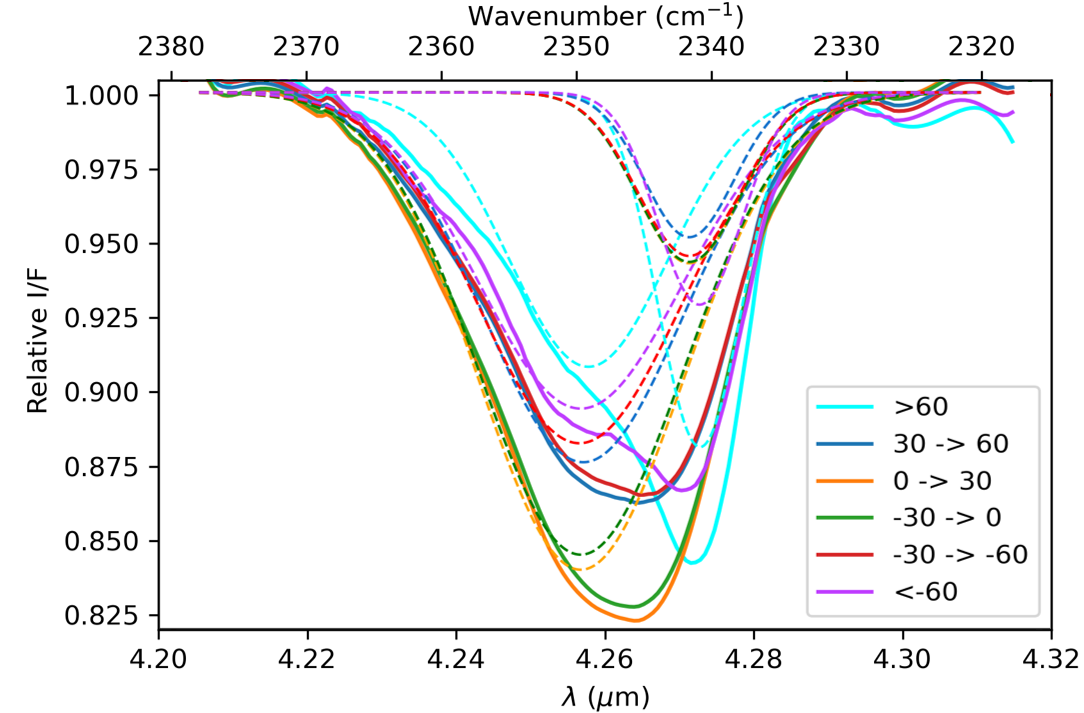}
    \caption{Latitude spectra of Ganymede’s leading hemisphere acquired with JWST. Colours are used to indicate different latitude ranges. The dotted lines represent the Gaussian fit of each individual spectrum which is deconvolved in two Gaussians.}
    \label{fig:Ganymede_leading+gaussian}
\end{figure}

Figure \ref{fig:Ganymede_leading+gaussian} shows the CO$_2$ band observed with JWST NIRspec on Ganymede as a function of  latitude. The solid lines are JWST spectra and the colours indicate different {latitudinal ranges, in good agreement with \cite{bockelee}}. The latitudinal variations of the CO$_2$ band and its asymmetric band shape support the view that different CO$_2$ ice physical states are involved. The  2342 cm$^{-1}$ (4.27 $\mu$m) peak is stronger across the northern latitudes, which are colder and more enriched in water ice \citep{bockelee,trumbo}. Across the equatorial latitudes, the  2353 cm$^{-1}$ (4.25 $\mu$m) band becomes stronger. Ganymede has an average surface temperature around 100 K, with a maximum of 160 K at the equator \citep{bockelee}. Each latitude spectrum can be deconvolved into two Gaussian distribution that are represented with dotted lines and plotted for each range of latitudes. When analyzing the results, we can see that one of the Gaussians is located at 2341.7 cm$^{-1}$ (4.270 $\mu$m) at the poles and shifts to lower wavelengths for other regions (supposedly warmer). This is in agreement with the $\tilde{\nu}_{3,3}$ band observed experimentally, attributed to CO$_2$ interacting with H$_2$O molecules in the ice, which shifts slightly to the blue as the temperature increases. Moving closer to the equator, this peak shifts even more to the blue, reaching its maximum at 2342.5 cm$^{-1}$ (4.269 $\mu$m). The second Gaussian is located around 2350.3 cm$^{-1}$ (4.255 $\mu$m) at the North pole and around 2350.9 cm$^{-1}$ (4.254 $\mu$m) at other latitudes. This is also in agreement with the $\tilde{\nu}_{3,1}$ band observed experimentally, which is blue-shifting from temperature above 150 K, which coincides with the surface temperature at Ganymede's equator. In line with the experimental results, the presence of the $\tilde{\nu}_{3,1}$ band is identified as CO$_2$ dangling bonds, which are an indication of ice porosity. 

\begin{figure}[h!]
    \centering
    \includegraphics[width=\hsize]{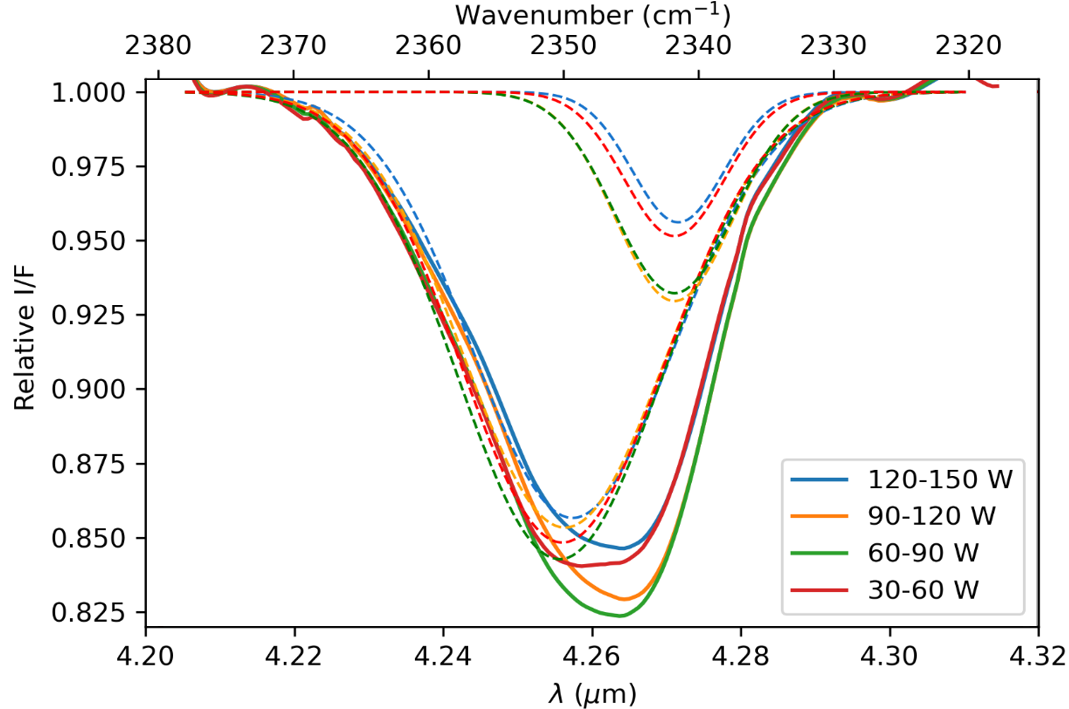}
    \caption{Longitudinal spectra of Ganymede’s leading hemisphere acquired with JWST. The longitudes are taken between 60$^{\circ}$ North and 60$^{\circ}$ South in latitude. The spectrum for 120-150 W represents the morning limb of the leading edge and the evening limb by the 30-60 W spectrum. Colors are used to indicate different longitudinal ranges. The dotted lines represent the Gaussian fit of each individual spectrum which is deconvolved into two Gaussians.}
    \label{fig:Ganymede_leading_LONG}
\end{figure}

The longitudinal spectra for Ganymede's leading hemisphere are shown in \autoref{fig:Ganymede_leading_LONG}. The solid lines are JWST spectra and the colours indicate different longitudinal ranges. The variations among the CO$_2$ band position and intensity are shown in the figure for latitudes between 60$^{\circ}$ north and 60$^{\circ}$ south. The morning limb is represented by the 120-150 W spectrum (blue), whereas the evening limb is represented by the 30-60 W spectrum (red). Each spectrum is deconvolved into two Gaussian distributions, which are shown by dotted lines. The band at 2351 cm$^{-1}$ (4.25 $\mu$m) has been assigned to the $\tilde{\nu}_{3,1}$ band being an indication of the CO$_2$ dangling bonds and of porosity. When comparing morning (blue) to evening (red), the $\tilde{\nu}_{3,1}$ band increases, suggesting that porosity increases during a day on Ganymede. This increase in porosity could be attributed to cracks in the ice that are forming due to temperature fluctuations. The effect of UV irradiation could also play an important role in this change in porosity that should also be considered. 

\subsection{Europa}
\begin{figure}[h!]
    \centering
    \includegraphics[width=\hsize]{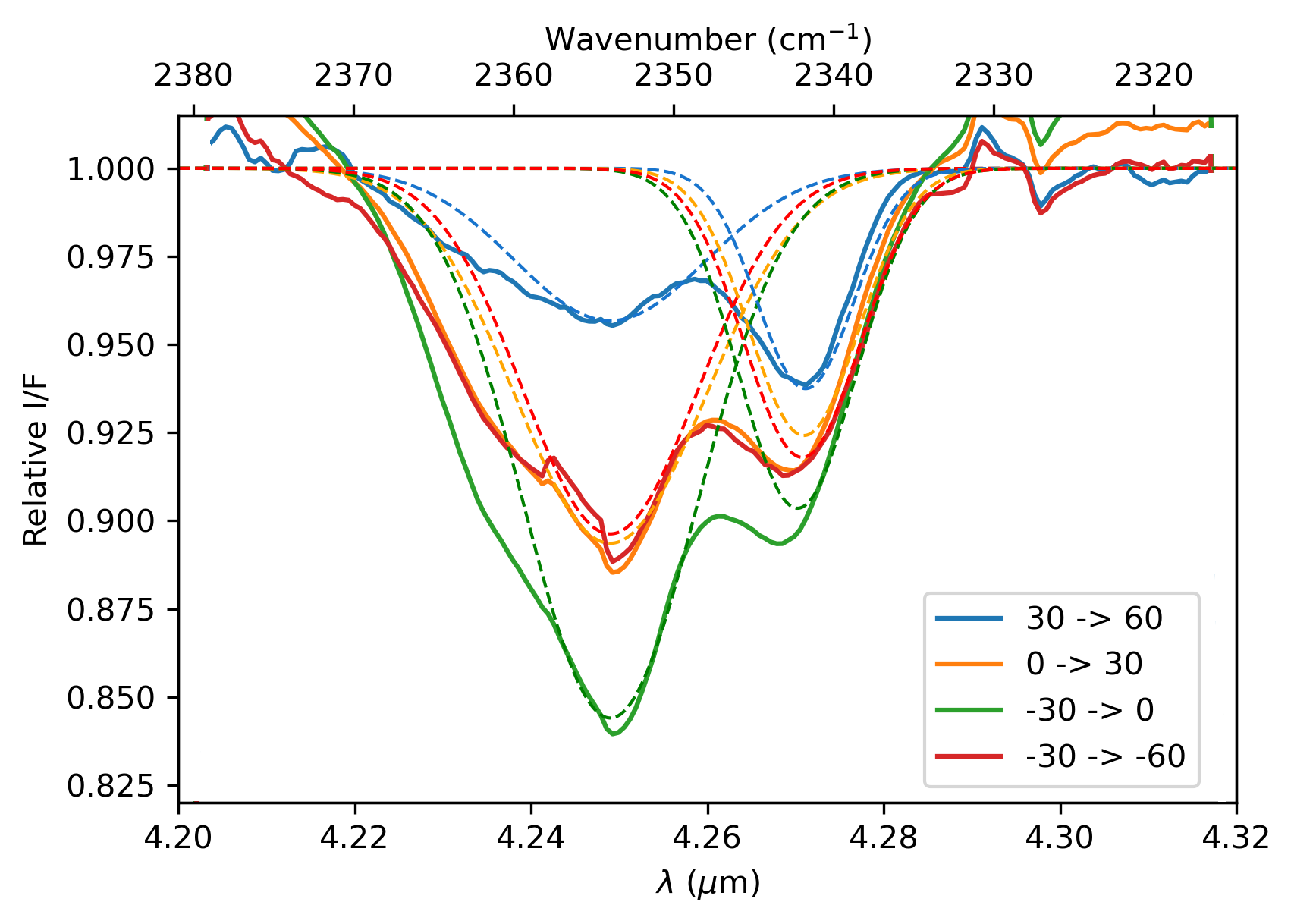}
    \caption{Latitude spectra of Europa’s leading hemisphere acquired with JWST. Colours are used to indicate different latitude ranges. The dotted lines represent the Gaussian fit of each individual spectrum, each consisting of two Gaussians.}
    \label{fig:europa+gaussian}
\end{figure}

Figure \ref{fig:europa+gaussian} shows the CO$_2$ band observed on Europa as a function of the latitude {and exhibits similar variation as that reported by \cite{villanueva}}. The band ca. 2342 cm$^{-1}$ (4.27 $\mu$m) is stronger across the northern latitudes, which are colder and more enriched in water ice (\citealt{trumbo2}). Across the equatorial latitudes, the 2353 cm$^{-1}$ (4.25 $\mu$m) peak becomes stronger. The temperature on Europa ranges from 60 K at the poles to around 110 K at the equator, which is colder than on Ganymede. The Gaussian fits of the data are represented with dotted lines, and plotted for each range of latitudes. Each latitude spectrum can be deconvolved into two Gaussian distributions (as seen in \autoref{fig:europa+gaussian}). The  band around 2342 cm$^{-1}$ (4.27 $\mu$m) feature shifts 
when moving closer to the equator, from 2340.9 cm$^{-1}$ (4.272 $\mu$m) for 30$^{\circ}$ to 60$^{\circ}$ North to 2341.2 cm$^{-1}$ (4.271 $\mu$m) at the equator. This is in agreement with the $\tilde{\nu}_{3,3}$ band position observed experimentally, which is located around 2340.8 cm$^{-1}$ (4.272 $\mu$m) and shifts to the blue with increasing temperatures. The second Gaussian $\tilde{\nu}_{3,1}$, located at 2353.5 cm$^{-1}$ (4.249 $\mu$m), is shifted to the blue by around 3 cm$^{-1}$ when compared to Ganymede. This is consistent with our experimental results showing the $\tilde{\nu}_{3,1}$ band shifts to higher wavelength with decreasing temperature. 

\noindent The main conclusion of this section is that CO$_2$ observations of both Ganymede and Europa can be explained by the presence of two components, the $\tilde{\nu}_{3,1}$ band and $\tilde{\nu}_{3,3}$ band. The $\tilde{\nu}_{3,3}$ band located at 2342 cm$^{-1}$ (4.27 $\mu$m) represents CO$_2$ trapped in water ice and can account for the CO$_2$ band observed in Ganymede's and Europa's spectra of the leading hemisphere, and is stronger in the boreal regions especially for Ganymede. The $\tilde{\nu}_{3,1}$ band at 2353 cm$^{-1}$ (4.25 $\mu$m) for Europa and at 2350.9 cm$^{-1}$ (4.254 $\mu$m) for Ganymede is stronger in other regions for both Ganymede and Europa. This band is associated with CO$_2$ dangling bonds which depend on the pores present in the ice, as CO$_2$ molecules on the pore walls can exhibit such dangling bonds. The band assignment of $\tilde{\nu}_{3,1}$ to CO$_2$ in pores has been suggested, based on our experimental data combined with DFT simulations. It should be noted that the  2353 cm$^{-1}$ (4.25 $\mu$m) band depends on the concentration of CO$_2$ in the water ice, which will make the band shift.  


\section{Conclusions}
In this study, we combined FTIR measurements and TPD experiments to explain the complexity of the CO$_2$ stretching band.
The examination of CO$_2$ ice behaviour on icy moons has offered intriguing revelations about its interaction within these distinct environments. Moreover, aligning laboratory findings and DFT calculations with JWST data from Ganymede and Europa, bridges the gap between theory, experiments, and celestial icy surfaces. These findings can help enrich our understanding of these moons' environments. 

\begin{itemize}
    \item The C--O stretching band observed for pure CO$_2$ ice at 10 K consists of two distinct vibrations: $\tilde{\nu}_{3,1}$ which is caused by porosity and peaks at 2351.3 cm$^{-1}$ (4.253 $\mu$m), and $\tilde{\nu}_{3,2}$ which corresponds to bulk CO$_2$ ice and peaks at 2345 cm$^{-1}$ (4.264 $\mu$m). The band's shape evolves during warm-up from 10 K to 80 K, exhibiting a redshift in both Gaussians. The TPD data indicates a desorption peak at 85 K, suggesting a thermal desorption of CO$_2$ from the ice layer. 
    \item CO$_2$ mixed with water ice (at low and high concentration) exhibits multiple peaks, of which two are the same as for pure CO$_2$. The presence of three peaks, $\tilde{\nu}_{3,1}$ at 2351 cm$^{-1}$ (4.25 $\mu$m), $\tilde{\nu}_{3,2}$ at 2345 cm$^{-1}$ (4.26 $\mu$m), and $\tilde{\nu}_{3,3}$ at 2341 cm$^{-1}$ (4.27 $\mu$m) at 10 K demonstrates the complexity of CO$_2$ interaction within water ice. The different CO$_2$ bands are assigned as follows: $\tilde{\nu}_{3,1}$ (2351 cm$^{-1}$, 4.25 $\mu$m) is associated to CO$_2$ dangling bonds in which CO$_2$ molecules are found in pores or cracks, $\tilde{\nu}_{3,2}$ (2345 cm$^{-1}$, 4.26 $\mu$m) is due to CO$_2$ trapped in the water ice (segregated) forming pockets of pure CO$_2$, and $\tilde{\nu}_{3,3}$ (2341 cm$^{-1}$, 4.27 $\mu$m) is related to CO$_2$ molecules embedded in the ice interacting with water molecules. 
    \item Thermal desorption analysis allowed for distinct peaks in the desorption curves to be assigned, indicating different desorption behaviours for CO$_2$ within the ice. The low and high concentrations display similar desorption patterns. Initial CO$_2$ desorption occurs around 80 K, followed by peaks at 146 K and 162 K, indicating various CO$_2$-water interactions during ice phase transitions. Continuous desorption between 80 and 146 K is observed, driven by the amorphous water ice rearrangement and thereby facilitating CO$_2$ diffusion through pores in the ice matrix.
    \item The JWST NIRSPEC spectra of Ganymede's leading hemisphere reveals significant variations in the CO$_2$ band profile across different latitudes. Two main CO$_2$ bands have been observed, that we assigned to $\tilde{\nu}_{3,1}$ (2353 cm$^{-1}$, 4.25 $\mu$m) and $\tilde{\nu}_{3,3}$ (2342 cm$^{-1}$, 4.27 $\mu$m). The dominance of $\tilde{\nu}_{3,1}$ (2353 cm$^{-1}$, 4.25 $\mu$m) in northern latitudes, associated with colder regions enriched in water ice, is in contrast with the prevalence of $\tilde{\nu}_{3,3}$ (2342 cm$^{-1}$, 4.27 $\mu$m) in equatorial latitudes. Gaussian fits of the spectra suggest two distinct physical states of CO$_2$ that shift with temperature, which confirm the assignments from our laboratory findings. The CO$_2$ band shifts observed on Ganymede with latitude could be attributed to an increase in temperature, showing that in the poles CO$_2$ is well embedded in water ice.
    \item The longitudinal spectra for Ganymede's leading hemisphere indicates variations in CO$_2$ spectra from morning to evening. The band is assigned to the $\tilde{\nu}_{3,1}$ band (2351 cm$^{-1}$, 4.25 $\mu$m) that links to the porosity of the ice. This increase in porosity could be attributed to cracks created in the ice due to variation of temperatures.
\end{itemize}    

    Similarly, the JWST NIRSPEC spectra of Europa's leading hemisphere can be deconvolved into two Gaussian fits: $\tilde{\nu}_{3,3}$ (2342 cm$^{-1}$, 4.27 $\mu$m) assigned to CO$_2$ molecules trapped in the water ice and $\tilde{\nu}_{3,1}$ (2353 cm$^{-1}$, 4.25 $\mu$m) assigned to dangling CO$_2$ molecules in pores or cracks. Temperatures in Europa are colder than Ganymede and as expected show a blueshift for $\tilde{\nu}_{3,1}$ and a redshift for $\tilde{\nu}_{3,3}$ when compared to the warmer moon. A similar effect is observed when comparing spectra from the warmer equator to the colder poles, indicating higher porosity near the equator.

\begin{acknowledgements}
The authors would like to thank I. de Pater, T. Fouchet, and D. Bockel\'ee-Morvan for the use of the JWST ERS Ganymede data {and assistance with processing}, as well as G. Villanueva for the use of the JWST Europa data. This research has been funded by project PID2020-118974GB-C21 by the Spanish Ministry of Science and Innovation. B.E. acknowledges support by grant PTA2020-018247-I by the Spanish Ministry of Science and Innovation/State Agency of Research MCIN/AEI.
\end{acknowledgements}

%
%

\bibliographystyle{aa}
\bibliography{biblio}

\end{document}